\documentclass[aps,
pra,
reprint,
]{revtex4-1}

\pdfoutput=1

\usepackage{natbib}
\setcitestyle{numbers}
\usepackage{graphicx}
\usepackage{xcolor}
\usepackage{amssymb, amsmath}
\usepackage{epstopdf}
\usepackage{units}
\definecolor{cadmiumgreen}{rgb}{0.0, 0.42, 0.24}

\usepackage[normalem]{ulem}

\begin{document}


\title{Emission of correlated  jets from a driven matter-wave soliton in a quasi-one-dimensional geometry}


\author{Tadej Me\v{z}nar\v{s}i\v{c}}
\email[]{tadej.meznarsic@ijs.si}
\author{Rok \v{Z}itko}
\author{Tina Arh}
\author{Katja Gosar}
\author{Erik Zupani\v{c}}
\author{Peter Jegli\v{c}}\email[]{peter.jeglic@ijs.si}
\affiliation{Jo\v{z}ef Stefan Institute, Jamova 39, SI-1000 Ljubljana, Slovenia}


\date{\today}

\begin{abstract}
We demonstrate the emission of correlated atom jets from a matter-wave soliton in a quasi-one-dimensional optical trap. We characterize the dependence of jet properties on the frequency, amplitude and length of the modulation, and qualitatively reproduce the trends in the mean-field picture with a one-dimensional time-dependent Gross-Pitaevskii equation simulation. High-order jets are observed for sufficiently long pulses. A double-pulse modulation sequence produces consecutive jets, and a multi-pulse sequence may lead to irregular 3D jets at a finite angle to the direction of the channel.  In the limit of vanishing high-order jets beyond-mean-field number correlations of jet pairs are demonstrated, implying possible number squeezing.
\end{abstract}


\maketitle


Periodic modulation of parameters of a physical system leads to novel phenomena.
Cold-atom systems offer excellent control of the trap geometry and interatomic interactions
and are therefore extremely suitable for such studies.
These range from resonantly exciting quadrupole modes in a trapped Bose-Einstein condensate (BEC) by modulating the interatomic interaction \cite{Pollack2010}, to parametrically amplifying the Faraday waves in a BEC by modulating the trapping potential \cite{Engels2007} or  the interaction \cite{Nguyen2019}, and creating synthetic gauge fields and various topological effects by modulating optical lattices \cite{Eckardt2017}.
Harmonic modulation of the interatomic interaction in a BEC stimulates collisions between the atoms and can lead to emission of matter-wave jets \cite{Clark2017}. 
The emission is preceded by the emergence of strong density waves in the condensate \cite{Fu2018}. 
Intricate angular correlations arise in the two-dimensional case \cite{Feng2019}.
The attempts at explaining the origin of the jets and their properties have employed a variety of theoretical approaches at different levels of sophistication with varying success \cite{Chen2018, Arratia2019, Wu2018}. 
However, the degree of correlation between jets and the possible utilization of jets for precision measurements remains largely unexplored.


In this Rapid Communication we present the observation of jets emitted from a self-trapped matter-wave soliton in quasi-one-dimensional (quasi-1D) geometry. 
First- and second-order jets are observed for a single modulation pulse and consecutive jets for a double modulation pulse.
For multiple modulation pulses irregular 3D jets appear, seemingly oblivious to the quasi-1D confinement.
All stages of the Bose jet emission are captured in a simple model based on the 1D Gross-Pitaevskii equation, giving an insight into the dynamics of density waves that precede the emission.
Asymptotically, the jets take the form of solitons with some radiation background.
Additionally, beyond the scope of the mean-field GPE simulation, the jet number correlations are investigated. 
We demonstrate sub-Poissonian statistics of the number difference between the left and right first-order jets in a regime of suppressed second-order jets, which could be an indication of number squeezing. 

We start by releasing a BEC of about 5000-10000 $^{133}$Cs atoms from a crossed dipole trap, 
into a channel with radial frequency $\omega_\mathrm{r} = 2\pi\cdot$\unit[101]{Hz} and a weak axial anti-trapping potential with "frequency" $2\pi\cdot$\unit[3.33]{Hz}. 
Via the broad \textit{s}-wave Feshbach resonance with zero crossing near \unit[17]{G} \cite{Chin2010} we tune the scattering length from positive (repulsive interaction) to slightly negative (attractive interaction) to obtain a bright matter-wave soliton, a nondispersing wave-packet that forms due to the nonlinearity of the interatomic interaction \cite{strecker2002formation, Khaykovich2002, Cornish2006, nguyen2014collisions,   McDonald2014, Lepoutre2016, nguyen2017formation, Everitt2017} (see Ref.~\onlinecite{Meznarsic2019} for further details of the procedure). 
Solitons maintain their density when released into the channel, ensuring stable conditions during the experiments. 


\begin{figure}[t]
\centering
\includegraphics[trim={0 1.5cm 0 3cm},width=1\linewidth]{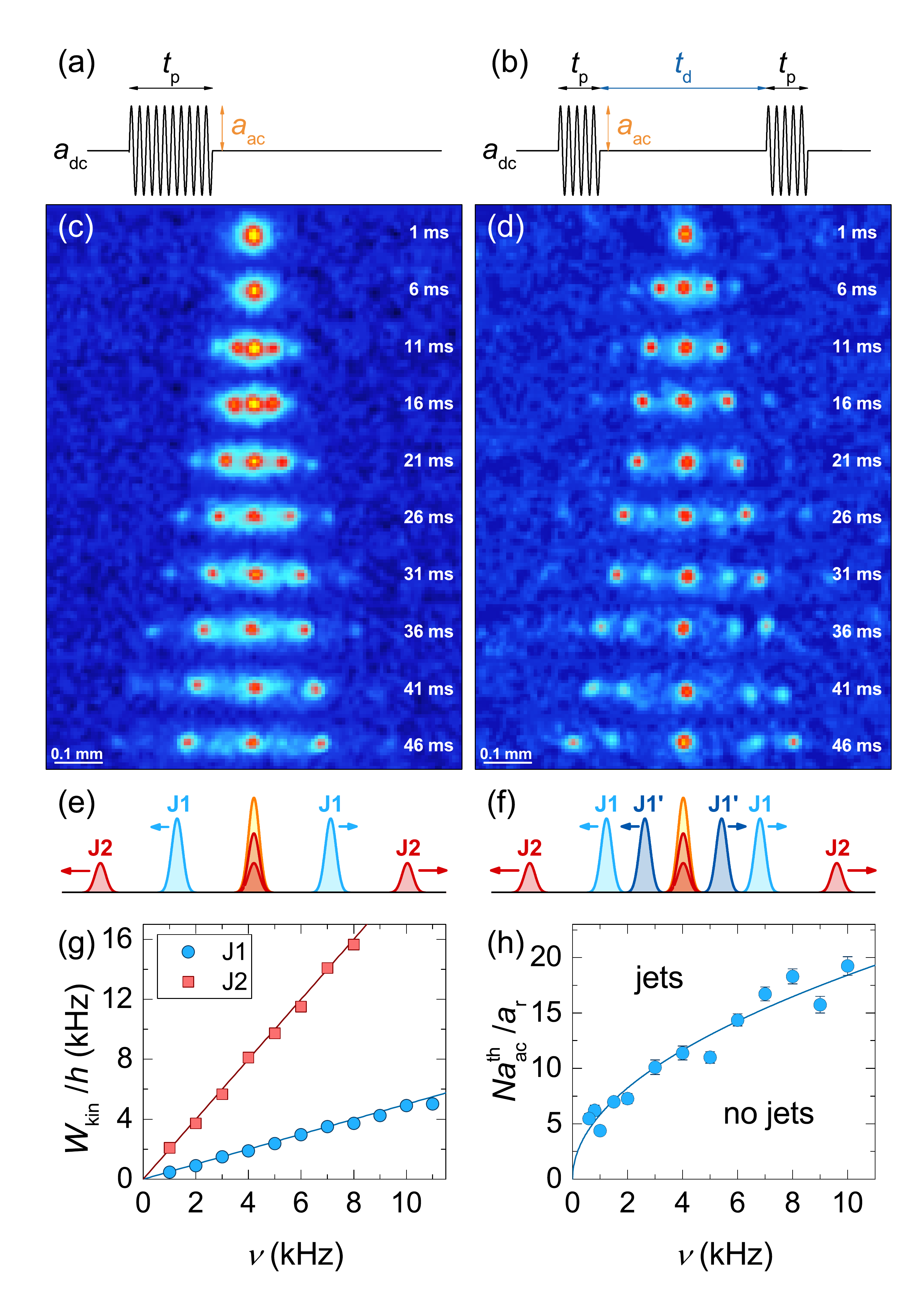}
\caption{\label{fig:1:Fireworks} 
(Color online) 
(a,b) Single- and double-pulse modulation sequences.  
The time evolution of jets (c) for \unit[2]{kHz}, $a_\mathrm{ac}=52a_0$, $t_\mathrm{p}=\;$\unit[11]{ms} single-pulse and (d) for \unit[4]{kHz}, $a_\mathrm{ac}=78a_0$, $\{t_\mathrm{p},\;t_\mathrm{d}\}=\;$\unit[\{6,\;10\}]{ms} double-pulse modulation sequence (\unit[15]{ms} time-of-flight). 
(e, f) Schematic representation of the jets in (c, d).
(g) Atom kinetic energy dependence on the modulation frequency for J1 and J2 jets. 
(h) Frequency dependence of the threshold modulation amplitude for jet formation ($t_\mathrm{p}=\;$\unit[50]{ms}).
All error bars indicate one standard error of the mean.  
}
\end{figure}

After releasing the BEC into the channel we modulate the scattering length for a finite time $t_\mathrm{p}$ as $a(t) = a_\mathrm{dc} + a_\mathrm{ac}\sin (2\pi\nu t)$, see Fig.~\ref{fig:1:Fireworks}(a), or we generate pulse-trains, see Fig.~\ref{fig:1:Fireworks}(b). 
Here, $a_\mathrm{ac}$ and $\nu$ are the amplitude and the frequency of the modulation, while $a_\mathrm{dc}$ is the solitonic scattering length.
In a typical experiment, modulation frequencies range from \unit[600]{Hz} to \unit[11]{kHz} with amplitudes up to 120$a_0$, where $a_0$ is the Bohr radius.
The modulation of the interaction triggers the emission of matter-wave jets from the soliton.
A typical time evolution for a single pulse is shown in Fig.~\ref{fig:1:Fireworks}(c).
The BEC symmetrically emits two pairs of jets with different velocities. 
The outer pair (``second-order'' jets, J2) is twice as fast as the inner pair (``first-order'' jets, J1).  
Fig.~\ref{fig:1:Fireworks}(g) shows the frequency dependence of the kinetic energy of the atoms in J1 and J2.
Atoms forming J1 have kinetic energy exactly $h\nu/2$, those in J2 exactly $2h\nu$, where $h$ is the Planck constant. 
From energy and momentum conservation it follows that J2 forms from the atoms in J1 rather than from the atoms in the central BEC, in fact, two J2 jets form from each J1, with half the atoms 
remaining inside the central cloud with zero velocity [see Fig.~\ref{fig:1:Fireworks}(e)].

In order for the jets to form, the amplitude of the modulation must exceed a threshold value of $a_\mathrm{ac}^\mathrm{th}$ \cite{Clark2017}.
The threshold [Fig.~\ref{fig:1:Fireworks}(h)] exhibits a square root dependence on the modulation frequency \cite{Clark2017}. 
Importantly, the interaction between the atoms depends not only on the scattering length but also on the density of atoms. 
Therefore, to account for atom number and thus density fluctuations  in different experimental runs, the interaction modulation in Fig.~\ref{fig:1:Fireworks}(h) is given as a dimensionless product $Na/a_\mathrm{r}$, where $N$ is the number of atoms in the soliton and $a_\mathrm{r} = \sqrt{\hbar/m \omega_\mathrm{r}}$ the harmonic oscillator length with reduced Planck constant $\hbar$ and atomic mass $m$.
In the double-pulse case we observe that an additional jet J1' is generated by the second pulse [Fig.~\ref{fig:1:Fireworks}(d,~f)]. 
It has the same initial velocity as the first J1. 
After the second pulse the condensate is too depleted and the threshold cannot be exceeded to create a third ``first-order'' jet.

\begin{figure}[t]
\centering
\includegraphics[trim={0 1cm 0 1.5cm},width=1\linewidth]{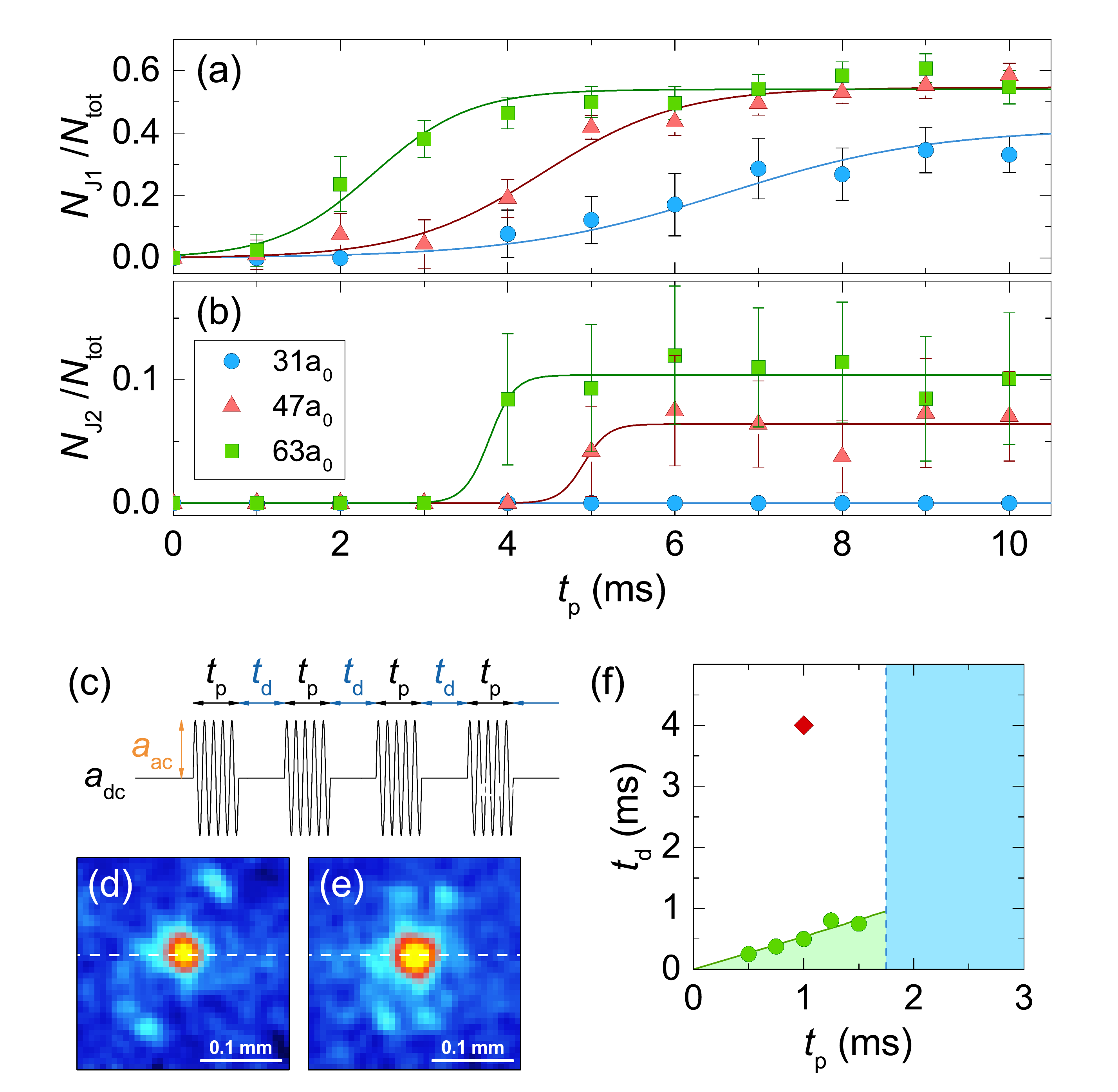}
\caption{\label{fig:2:Fireworks} 
(Color online) 
(a, b) Fraction of atoms in J1 and J2 jets as a function of pulse length for \unit[4]{kHz} modulation of different amplitudes. 
The solid lines are $b/[c+\exp ( -a\cdot t_\mathrm{p})]$ fits to the data. 
(c) Multi-pulse train used in the experiments shown in (d, e). 
(d, e) Irregular 3D jets formed at a finite angle to the direction of the channel (white dashed line), for $t_\mathrm{p} = \;$\unit[1]{ms} and $t_\mathrm{d} = \;$\unit[4]{ms}, marked by the red diamond in (f). 
(f) Phase diagram for jet formation with \unit[4]{kHz}, $a_\mathrm{ac}=47a_0$ pulse trains.
All error bars indicate one standard error of the mean.
}
\end{figure}

For $a_\mathrm{ac}$ above the threshold, the number of atoms in the jets as a function of the pulse duration $t_\mathrm{p}$ first exponentially increases and then slowly saturates.
The exponential growth is characteristic for bosonic stimulation \cite{Miesner1998}, while the saturation happens due to the depletion of the central BEC:
with decreasing number of atoms in the cloud the frequency of collisions decreases and the jet formation processes gradually become rarer until they completely stop.
The ratio between the number of atoms in the fully-formed jets and the total number of atoms is shown for different modulation amplitudes $a_\mathrm{ac}$ in Fig.~\ref{fig:2:Fireworks}(a,~b) for J1 and J2, respectively. 

Figs.~\ref{fig:2:Fireworks}(a,~b) indicate that jets cannot form if the modulation pulse is too short. 
Nevertheless, it is possible to generate jets by exciting the BEC with a train of several short pulses of length $t_\mathrm{p}$ separated by time delays $t_\mathrm{d}$, see Fig.~\ref{fig:2:Fireworks}(c). 
The total duration of the pulse train is fixed, meaning that there are more pulses in trains with shorter $t_\mathrm{p}$ and $t_\mathrm{d}$.
$t_\mathrm{p}$ and $t_\mathrm{d}$ are always a multiple of the modulation pulse oscillation period to avoid destructive interference effects.
From the results one can establish a phase diagram Fig.~\ref{fig:2:Fireworks}(f). 
In the blue (dark gray) region ($t_\mathrm{p}>\;$\unit[1.75]{ms}) the jets are emitted after a single pulse and in the green (light gray) region after multiple pulses.


\begin{figure}[t]
\centering
\includegraphics[width=1\linewidth]{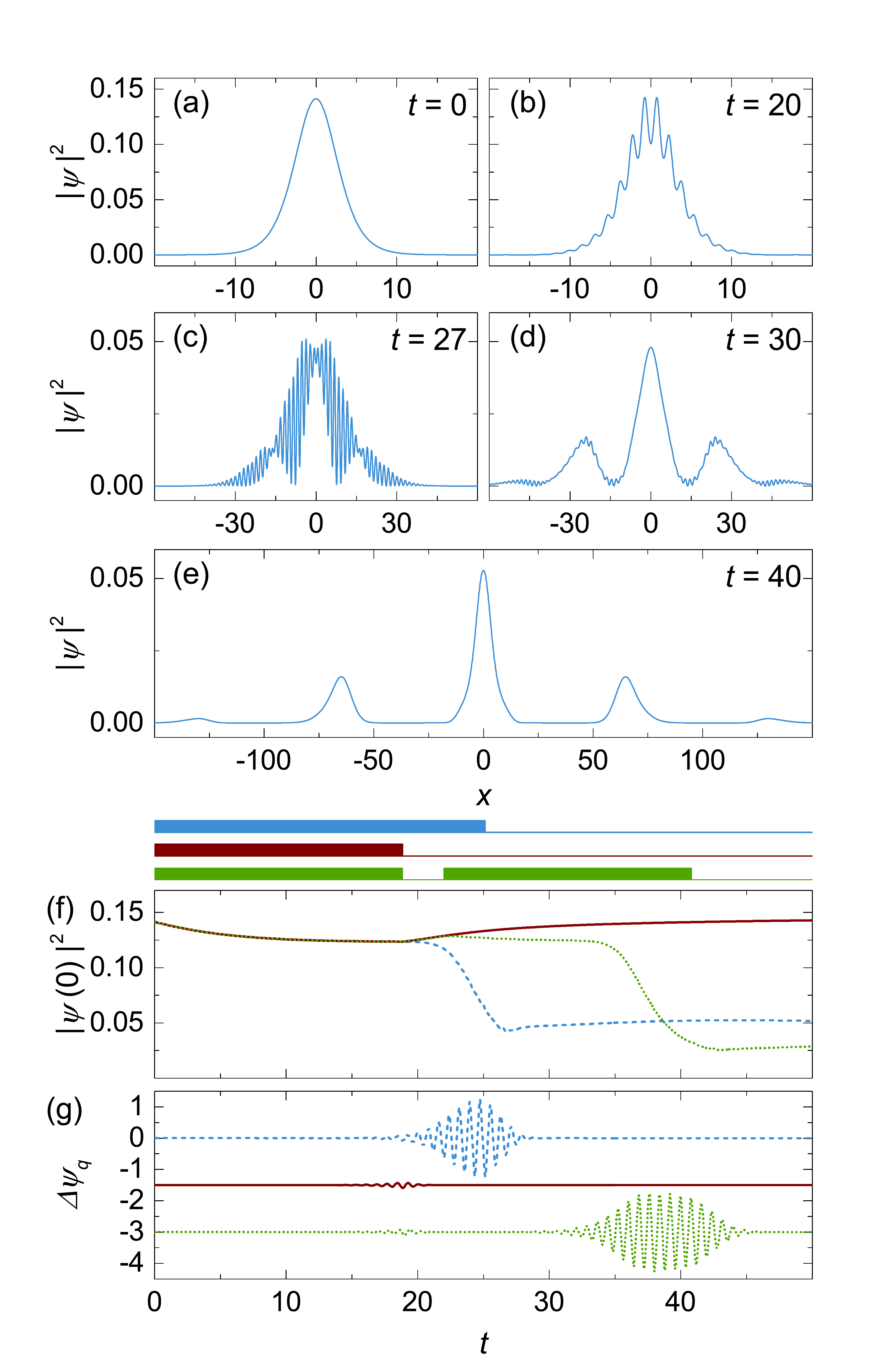}
\caption{\label{fig:3:Fireworks} 
(Color online)
Simulation. (a-e) Snapshots of condensate density during the formation of jets.
Please note the changing scales in different panels.
The model parameters are $k_\mathrm{dc}=-0.045$, $k_\mathrm{ac}=1.8$, $\omega=16$, $t_\mathrm{p}=8\pi$.
Time-dependence of the central soliton amplitude (f) and density modulation amplitude (g) for the modulation pulse sequences shown above (f) (blue/upper/dashed: $t_\mathrm{p}=8\pi$, red/middle/solid: $t_\mathrm{p}=6\pi$ and green/lower/dotted:  $t_\mathrm{p}=6\pi,\;t_\mathrm{d}=\pi$).
The solid red and dotted green curve in (g) are offset for clarity. 
}
\end{figure}

In the white region  jets along the direction of the channel do not appear. However, in rare cases jets at a finite angle to the channel are observed, as shown in Figs.~\ref{fig:2:Fireworks}(d,~e).
The angle of these irregular 3D jets appears to be random, and multiple jets can also be observed as shown in Fig.~\ref{fig:2:Fireworks}(e). 
They have insufficient energy to escape the confinement potential, thus they oscillate in the channel and can only be observed after an appropriately long time-of-flight (usually \unit[15]{ms}).

Microscopic processes responsible for the ejection of atoms have been identified to be photon-stimulated two-atom collisions \cite{Clark2017}. 
This can be modeled through Bogoliubov approximation by separating the field operator into
negligibly-depleted condensate and an excited-mode field \cite{Clark2017}, through more involved methods \cite{Wu2018,Chen2018}, or by numerically solving the time-dependent Gross-Pitaevskii equation (GPE) \cite{Fu2018,Salasnich2003}. 
Here we used the latter approach in 1D. 
We find that 
the formation of the density wave and its exponential growth, the emission of the jets
(including the formation of the higher-order jets J2), the  frequency dependences of density-wave wavelength and kinetic energy of the jets, the existence of various thresholds, as well as the qualitative functional forms of all dependences shown in Figs.~\ref{fig:1:Fireworks},~\ref{fig:2:Fireworks} are captured correctly in this simple picture.
However, quantum number correlations discussed later in the paper go beyond the scope of the mean-field approach and cannot be reproduced with GPE simulations.

Fig.~\ref{fig:3:Fireworks}(a-e) shows snapshots of the condensate density $n=|\psi(x)|^2$ for a single pulse modulation sequence at five moments: (a) initial state, (b) emerging density wave, (c) strongly perturbed condensate at the moment of jet ejection, (d) resulting J1 and J2 jets showing residual
density modulation, (e) asymptotic state with one stationary soliton and two pairs of traveling solitons. 
We also show the time-dependence of the amplitude of the central soliton in Fig.~\ref{fig:3:Fireworks}(f) and the amplitude of the density modulation in Fig.~\ref{fig:3:Fireworks}(g) for three different modulation sequences.
Interatomic interaction is given as a dimensionless parameter $k=Na/a_\mathrm{r}$, modulation frequency as $\omega=2\pi\nu$, time in units $1/\omega_\mathrm{r}$ and distances in units $a_\mathrm{r}$.
In the early stages, the BEC slightly changes its shape due to the rectified effect of the modulation [visible as the relaxation of the soliton amplitude up to $t \approx 12$ in panel (f)], and hardly perceptibly expands and contracts as a whole, i.e., the breathing mode is being excited. The density modulation with wave number $q = \sqrt{m\omega/\hbar}$ becomes appreciable for $t \gtrsim 15$. 
The jets start to form at $t \approx 25$ [dashed blue in (f, g)]. 
The density wave amplitude starts to decay when the modulation pulse ends 
at $t_\mathrm{p}=8\pi$ and eventually the waveforms of both the central peak and the jets J1 and J2 tend toward smooth soliton-like lineshapes (but never reaching the ideal $1/\cosh(x)$ form). 
If the driving modulation ends prematurely, the density wave disappears, no jets are emitted and the BEC returns to the initial solitonic shape [solid red in (f, g)].
However, a second driving pulse after a sufficiently short delay can revive the almost extinguished density wave leading to jet emission [dotted green in (f, g)].
These examples give further insight into the boundary between the green (light gray) and white region in Fig.~\ref{fig:2:Fireworks}(f).
If the delay $t_\mathrm{d}$ between the modulation pulses is short, so that the density wave decays only partially, jets are generated [green (light gray) region],  otherwise they are not (white region).
For longer pulse times $t_\mathrm{p}$ the density wave amplitude is higher and takes longer to decay. 
The boundary between the two regions is linear, because growth (during time $t_\mathrm{p}$) and decay (during time $t_\mathrm{d}$) of the modulation both have exponential time dependence.
The slope of the boundary can be written as $t_\mathrm{d}/t_\mathrm{p} = A/B-1$, where  $A$ is the growth rate and $B$ the decay rate. 
At the threshold for jet formation the two rates are equal ($A/B=1$) and the green (light gray) region  vanishes. 
The slope in Fig.~\ref{fig:2:Fireworks}(f) is 0.54(3), which matches the ratio 1.54(14) between the modulation amplitude for this measurement and the threshold measurement at \unit[4]{kHz} [Fig.~\ref{fig:2:Fireworks}(b)].

\begin{figure}[t]
\centering
\includegraphics[width=1\linewidth]{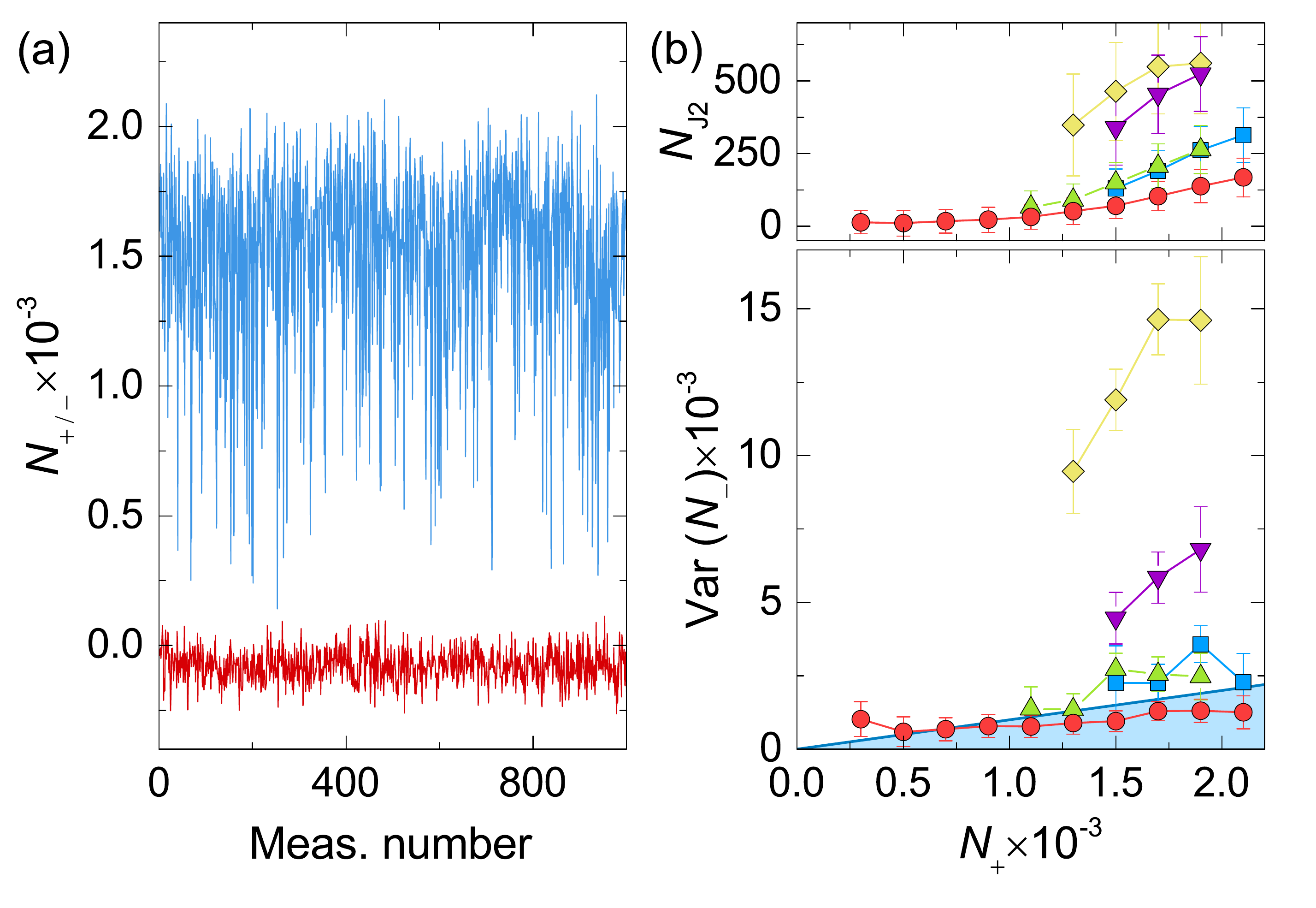}
\caption{\label{fig:4:Fireworks}
(Color online) 
(a) Sum $N_+$ (blue/light gray) and difference $N_-$ (red/dark gray) of the atom numbers in the left and right  J1 jets in a typical series of measurements.
(b) Lower panel: variance of $N_-$ with subtracted detection noise for a series of measurements with different modulation pulse durations, frequencies and amplitudes \cite{Note1}. Number of atoms in J2 jets is shown in the upper panel. The blue shaded area indicates the sub-Poissonian regime.
}
\end{figure}

Because of the momentum conservation during the jet emission process one naturally expects the same number of atoms in the left and right jet ($N_\mathrm{L}$ and $N_\mathrm{R}$). 
Random processes such as interatomic collisions that produce matter-wave jets exhibit Poissonian statistics, which means that the variance of the number of atoms in either jet should be the same as its average over many measurements $\langle N_\mathrm{L}\rangle$ ($\langle N_\mathrm{R}\rangle$).
If the left and right jet were created independently, the sum $N_+=N_\mathrm{L}+N_\mathrm{R}$ and difference $N_-=N_\mathrm{L}-N_\mathrm{R}$ would also have Poissonian distributions with variance $\langle N_+\rangle$.
But because the jets form with pairwise collisions of condensate atoms the number difference is no longer random and its variance should be below the shot-noise (sub-Poissonian).
Accordingly, the variable $N_+$ should have a variance larger than shot-noise (super-Poissonian).
Sub-Poissonian fluctuations are a prerequisite for many-body entanglement \cite{Buecker2011, Gross2011, Lucke2011, Bonneau2013, Bookjans2011}.

Fig.~\ref{fig:4:Fireworks}(a) shows the number difference between the left and right J1, $N_-$, and the total number in J1, $N_+$, over 1000 measurements.
In order to suppress the effects of variable atom number 
the measurements are binned according to the total number $N_+$, and the variance of the difference $\mathrm{Var}(N_-)$ is calculated for each bin (chosen bin size $\Delta_\mathrm{bin}=200$).
The measured variance is larger than the actual variance due to detection noise, which we determine from an empty part of the image \footnote{See Supplemental Materials}.
To determine if the distribution of $N_-$ is sub-Poissonian, we subtract the detection noise and compare it to the expected Poissonian variance $\langle N_+\rangle$ [Fig.~\ref{fig:4:Fireworks}(b)].
From the measurements shown in the Fig.~\ref{fig:4:Fireworks}(b) we can see that $\mathrm{Var}(N_-)$ is strongly dependent on the number of atoms in the second order jet J2.
For larger $N_\mathrm{J2}$ the variance is larger.
This is caused by the asymmetric formation of J2 jets from the left and right J1 jet.
It is therefore very important to choose a suitable pulse length and modulation amplitude to reduce $N_\mathrm{J2}$ as much as possible while maintaining large $N_\mathrm{J1}$ (See Fig.~\ref{fig:2:Fireworks}(a,~b) and Supplemental Materials \cite{Note1}).
Only with small enough $N_\mathrm{J2}$ can $\mathrm{Var}(N_-)$ reach the sub-Poissonian regime [red circles in Fig.~\ref{fig:4:Fireworks}(b)], which implies possible number squeezing and makes further studies of many-body entanglement possible  \cite{Fadel2018, Kunkel413, Lange2018}.




In conclusion, our experiment demonstrates the emission of matter-wave jets from a self-trapped BEC (soliton) due to the modulation of interatomic interaction in quasi-1D confinement.
While in a single-pulse experiment only the first and second order jets are created, the double-pulse experiment creates two consecutive first order jets, implying a possibility of multiple consecutive jets.
With a high enough number of atoms in the initial condensate one could in principle perform the experiment shown in Fig.~\ref{fig:1:Fireworks}(b, d) with more than two pulses and thereby create a pulsed atom laser with two correlated beams, a powerful new tool for precision measurements.
The creation of symmetric correlated jets would also be an interesting alternative to light-pulse beam splitters \cite{Angelis2008}.
The velocity of jets can be tuned continuously with the modulation frequency in contrast to the quantized light momentum imprinted through the excitation of an atomic transition. 
In multi-pulse experiments new, irregular 3D jets emerge, indicating that in some cases the confinement does not define the direction of the jets, which merits further investigation.
Furthermore, we show that a relatively simple numerical calculation predicts the qualitative behaviour of the jets, barring the number correlations, and additionally shows the emergence of density waves before jet emission.
The simplified 1D geometry removes the angular complexity of previous 2D experiments, making further analysis more straightforward, as exemplified by the number correlation measurements.
The degree of first-order jet correlations depends on the number of atoms in the second-order jets, which can be controlled in our experiment. The precise control of second-order jets could also be relevant for quantum simulations of Unruh thermal radiation recently demonstrated in 2D geometry \cite{Hu2019}.
It would also be interesting to study the effects of modulation on higher order solitons, recently demonstrated with a cesium BEC \cite{Carli2019}, where in addition to emission of jets one expects the soliton to split into its constituent fundamental solitons \cite{Sakaguchi2004,Yanay2009}.


We thank Antun Bala{\v z}, Naceur Gaaloul, Philipp Haslinger, Boris Malomed, Stephanie Manz, Marcos Rigol, Andrea Trombettoni, Lev Vidmar and Andrej Zorko for their comments. 
We would also like to thank Samo Begu{\v s} and Davorin Kotnik for their help with electronics.
This work was supported by the Slovenian Research Agency (research core Grants No. P1-0125 and No. P1-0099, and research project No. J2-8191).

\bibliographystyle{apsrev4-1}
%

\end{document}